# Generation of Arbitrarily Multiple Entangled Fields via Mechanical Oscillator Displacement

Xihua Yang[1], Junxiang Zhang[2], Jingping Xu[3], and Ligang Wang[2]

[1]*Department of Physics, Shanghai University, Shanghai 200444, China*

[2]*Zhejiang Province Key Laboratory of Quantum Technology and Device, Department of Physics, Zhejiang University, Hangzhou 310027, China*

[3]*Key Laboratory of Advanced Micro-Structured Materials of Ministry of Education, School of Physics Science and Engineering, Tongji University, Shanghai 200092, China*

We present a convenient and efficient scheme to generate arbitrarily multipartite continuous-variable entanglement via mechanical oscillator displacement induced by two strong input pump fields in the conventional single-cavity optomechanical system. It is shown that multipartite entanglement among the outputs of the two pump fields and any number of relatively weak probe fields can be realized and optimized when the two pump fields with suitable amplitude ratio and relative initial phase are, respectively, tuned to the red and blue mechanical sidebands of a single cavity mode and each probe field is red-detuned by the mechanical frequency with respect to a different neighboring cavity mode. This method can, in principle, be extended to flexibly and conveniently generate arbitrarily multiple nondegenerate bright entangled fields by using only coherent laser fields, and may find promising applications in realistic quantum communication and networks.

Generation of multipartite continuous-variable (CV) entanglement plays an essential role in quantum communication and networks [1, 2]. As is well known, realistic quantum networks would be composed of many quantum nodes and channels, where at the quantum nodes, multiple nondegenerate bright entangled fields are quite necessary for connecting different physical systems, and the atomic ensembles [3-9] or mechanical oscillators [10-30] with long coherence time provide potential matter media for storage and manipulation of quantum information. Therefore, how to conveniently and efficiently produce high degree of light-light, light-matter, and matter-matter multipartite entanglement remains a challenging task in the process of realizing realistic quantum communication and networks.

To compare with the atomic system, the optomechanical system can, in principle, couple to light fields with any desired frequencies for producing multiple entangled fields (e.g., the generation of bipartite entanglement between microwave photon and optical photon [11, 12]), thereby avoiding the enquire of particular frequencies corresponding to naturally existing atomic resonances in the atomic system; moreover, the optomechanical system can also be employed to test quantum features of macroscopic mechanical objects, which would be beneficial to investigating directly the fundamentals of quantum mechanics and the limitations of quantum-based measurements [13, 14]. Many schemes have been proposed for generating entanglements between a cavity field mode and a mechanical mode, between two or more cavity field modes as well as mechanical modes, such as driving one single sideband, two sidebands of a single cavity mode, and two or more independent cavity modes with one or more laser fields [15-24]. In addition, several kinds of driving methods, such as squeezed input laser fields [25, 26], classical feedback based on the processing of outcomes [27, 28], and time-periodically modulated coherent driving fields [29, 30], have been utilized to dramatically improve the degree of quantum squeezing and entanglement in the optomechanical system.

The strong interactions between laser fields and material media can result in many special effects, such as electromagnetically induced transparency (EIT) [31-33], electromagnetically induced entanglement (EIE) [34, 35], and electromagnetically induced squeezing (EIS) [36, 37] in the atomic system, where the common physical mechanism is the atomic coherence induced by laser fields. Recently, the optomechanical analog of EIT, EIE, and EIS, that is, optomechanically induced transparency, entanglement, and squeezing (OMIT [38-40], OMIE [41], and OMIS [18-20]), have also been extensively studied, where the physical origin of these peculiar effects is mechanical oscillator displacement, which plays a role similar to the atomic coherence for EIT, EIE, and EIS in the atomic system.

Motivated by the proposal of generating multipartite CV entanglement via atomic spin coherence created by the strong on-resonant coupling and probe fields in the Λ-type EIT-configuration atomic system [8], in this study, we present a convenient and efficient scheme to generate arbitrarily multiple nondegenerate entangled fields via mechanical oscillator displacement induced by two strong input pump fields in the conventional single-cavity optomechanical system. The two strong pump fields with suitable amplitude ratio and relative initial phase are tuned, respectively, to the red and blue mechanical sidebands of a single cavity mode to produce mechanical oscillator displacement, which acts as a quantum entanglement mediator, realizing, in principle, multipartite entanglement among the outputs of the two strong pump fields and any number of relatively weak probe fields with each probe field red-detuned by the mechanical frequency to a different neighboring cavity mode. This scheme provides an efficient and convenient way to generate multiple nondegenerate bright entangled light beams by using only coherent laser fields, and may bring great facility in realistic quantum communication and networks.

**Theoretical model and Heisenberg-Langevin equations.** The considered standard single-cavity optomechanical system is driven by two strong input pump 1 and 2 fields with frequencies $\omega_{l1}$ and $\omega_{l2}$ and any number of relatively weak input probe fields 1, 2,…n (n being a positive integer) with frequencies $\omega_{p1,p2,...pn}$ (see Fig. 1a), and the relevant frequencies of the cavity fields as well as the input pump and probe fields are displayed in Fig. 1b, where the input pump 1 (2) field is red (blue)-detuned by the mechanical oscillation frequency $\omega_m$ to the cavity field mode 0 with frequency $\omega_{c0}$, and the input probe fields are tuned to the red mechanical sidebands of the neighboring cavity modes with frequencies $\omega_{c1,c2,...cn}$. The Hamiltonian of the cavity optomechanical system can be described as [29, 39]

$$H = \sum_{i=0}^{n}\hbar\omega_{ci}a_i^+a_i + \hbar\omega_m b^+b - \sum_{i=0}^{n}\hbar g_i a_i^+ a_i(b^+ + b) + (i\hbar\eta_{l1}a_0^+ e^{-i\omega_{l1}t} + i\hbar\eta_{l2}a_0^+ e^{-i(\omega_{l2}t-\varphi)} + \sum_{j=1}^{n}i\hbar\eta_{pj}a_j^+ e^{-i\omega_p t} + h.c.), \quad (1)$$

where $a_i$ ($b$) is the annihilation operator of the cavity field mode i (the mechanical oscillation mode) and $g_{0(j)} = \omega_{l1,l2(pj)}\sqrt{\hbar/m\omega_m}/L$ is the optomechanical coupling coefficient of the radiation pressure with $m$ being the effective mass of the mechanical oscillator and $L$ the cavity length. The last three terms in Eq. (1) describe the interaction of the two pump and n probe input fields with the cavity field modes, where $\eta_{l1,l2(pj)}$ is related to the input pump (probe) field power $P_{l1,l2(pj)}$ with $\eta_{l1,l2(pj)} = \sqrt{2P_{l1,l2(pj)}k/\hbar\omega_{c0(j)}}$ (for simplicity, we assume all of the decay rates of the cavity fields are equal to $k$), and $\varphi$ is the relative initial phase between the two input pump fields. We denote the frequency detuning of the cavity mode 0 (input pump 2 field) with respect to the input pump 1 field as $\Delta_0 = \omega_{c0} - \omega_{l1}$ ($\delta = \omega_{l2} - \omega_{l1}$), and the frequency detuning of the cavity mode j to the corresponding input probe field as $\Delta_j = \omega_{cj} - \omega_{pj}$. In the frame rotating at the input pump 1 (probe j) field frequency $\omega_{l1}$ ($\omega_{pj}$), the Heisenberg-Langevin equations can be written as

$$\dot{a}_0 = -(k+i\Delta_0)a_0 + ig_0 a_0(b+b^+) + \eta_{l1} + \eta_{l2}e^{i\varphi}e^{-i\delta t} + \sqrt{2k}a_0^{in}, \quad (2a)$$

$$\dot{b} = -(\gamma_m + i\omega_m)b + \sum_{i=0}^{n}ig_i a_i^+ a_i + \sqrt{2\gamma_m}b^{in}, \quad (2b)$$

$$\dot{a}_j = -(k+i\Delta_j)a + ig_j a_j(b+b^+) + \eta_{pj} + \sqrt{2k}a_j^{in}, \quad (2c)$$

where $\gamma_m$ is the damping rate of the mechanical oscillator, and $a_i^{in}(t)$ and $b^{in}(t)$ are the optical and mechanical noise operators with the relevant nonzero correlation functions $\langle a_i^{in}(t)a_i^{in+}(t')\rangle = \delta(t-t')$ and $\langle b^{in}(t)b^{in+}(t')\rangle = (N+1)\delta(t-t')$ in the limit of large mechanical quality factor (i.e., $Q_m = \omega_m / \gamma_m \gg 1$) with $N = 1/(\exp(\hbar\omega_m/k_B T)-1)$ being the mean thermal phonon number, $k_B$ the Boltzmann constant, and $T$ the mirror temperature. We further assume that the strengths of the two pump fields are far larger than that of the probe field ($P_{l1,l2} \gg P_{pj}$), so the mechanical oscillator displacement mainly results from the two pump fields, and the radiation pressure force from the probe fields can be safely neglected. Consequently, in the resolved sideband regime ($\omega_m \gg \kappa$), solutions to Eqs. (2a-2b) can be well approximated by the ansatz $a_0 = a_{0^-} + a_{0^+} e^{-i\delta t}$ and $b \doteq b_0$, where $a_{0^-}$ and $a_{0^+}$ correspond, respectively, to the cavity field operators with frequency components of the pump 1 ($\omega_{l1}$) and pump 2 ($\omega_{l2}$) fields in the original frame. Similar treatment has been employed to produce mechanical squeezing in an electromechanical system [27] and OMIE in an optomechanical system [41]. Submitting $a_0$ and $b$ into Eqs. (2a-2b) and equating the respective frequency components, the evolutions of the operators $a_{0^-}$, $a_{0^+}$, $a_j$, and $b_0$ can be written as

$$\dot{a}_{0^-,j} = -(k+i\Delta_{0,j})a_{0^-,j} + ig_{0,j}a_{0^-,j}(b_0 + b_0^+) + \eta_{l1,pj} + \sqrt{2k}a_{0^-,j}^{in}, \tag{3a}$$

$$\dot{a}_{0^+} = -(k+i(\Delta_0 - \delta))a_{0^+} + ig_0 a_{0^+}(b_0 + b_0^+) + \eta_{l2}e^{i\varphi} + \sqrt{2k}a_{0^+}^{in}, \tag{3b}$$

$$\dot{b}_0 = -(\gamma_m + i\omega_m)b_0 + ig_i \sum_{i=0}^{n} a_i^+ a_i + \sqrt{2\gamma_m}b_0^{in}. \tag{3c}$$

By writing each Heisenberg operator in Eqs. (3a-3c) as the sum of its steady-state mean value and a small fluctuation operator with zero-mean value $a_s = \alpha_s + \delta a_s$ ($s = 0^-, 0^+, j$), and $b_0 = \beta + \delta b$, one can readily get the steady state mean values $\alpha_{0^-,j} = \eta_{l1,pj}/|k+i[\Delta_{0,j} - g_{0,j}(\beta+\beta^*)]|$, $\alpha_{0^+} = e^{i\varphi}\eta_{l2}/(k+i(\Delta_0 - \delta - g_0(\beta+\beta^*)))$, and $\beta \doteq ig_0(|\alpha_{0^-}|^2 + |\alpha_{0^+}|^2)/(\gamma_m + i\omega_m)$, where the phase references of the input pump 1 field and input probe fields are chosen to let $\alpha_{0^-,j}$ be real and positive. Defining the fluctuation quadrature operators $\delta X_s=(\delta a_s+\delta a_s^+)/\sqrt{2}$ and $\delta Y_s=(\delta a_s-\delta a_s^+)/\sqrt{2}i$, and $\delta X_b=(\delta b+\delta b^+)/\sqrt{2}$ and $\delta Y_b=(\delta b-\delta b^+)/\sqrt{2}i$, and the corresponding Hermitian noise operators $X_s^{in}$ and $Y_s^{in}$, as well as $X_b^{in}$ and $Y_b^{in}$, we can obtain the quantum Langevin equations for the fluctuation operators

$$\delta \dot{X}_{0^-,j} = -k\delta X_{0^-,j} + [\Delta_{0,j} - g_{0,j}(\beta+\beta^*)]\delta Y_{0^-,j} + \sqrt{2k}X_{0^-,j}^{in}, \tag{4a}$$

$$\delta \dot{Y}_{0^-,j} = -k\delta Y_{0^-,j} - [\Delta_{0,j} - g_{0,j}(\beta+\beta^*)]\delta X_{0^-,j} + 2g_{0,j}\text{Re}(\alpha_{0^-,j})\delta X_b + \sqrt{2k}Y_{0^-,j}^{in}, \tag{4b}$$

$$\delta \dot{X}_{0^+} = -k\delta X_{0^+} + [\Delta_0 - \delta - g_0(\beta+\beta^*)]\delta Y_{0^+} - 2g_0\text{Im}\alpha_{0^+}\delta X_b + \sqrt{2k}X_{0^+}^{in}, \tag{4c}$$

$$\delta \dot{Y}_{0^+} = -k\delta Y_{0^+} - [\Delta_0 - \delta - g_0(\beta+\beta^*)]\delta X_{0^+} + 2g_0\text{Re}\alpha_{0^+}\delta X_b + \sqrt{2k}Y_{0^+}^{in}, \tag{4d}$$

$$\delta \dot{X}_b = -\gamma_m \delta X_b + \omega_m \delta Y_b + \sqrt{2\gamma_m}X_b^{in}, \tag{4e}$$

$$\delta \dot{Y}_b = -\gamma_m \delta Y_b - \omega_m \delta X_b + 2g_0 \sum_{s=0^-,0^+,1}^n \text{Re}(\alpha_s)\delta X_s + 2g_0 \sum_{s=0^-,0^+,1}^n \text{Im}(\alpha_s)\delta Y_s + \sqrt{2\gamma_m}Y_b^{in}. \tag{4f}$$

By Fourier-transforming Eqs. (4a-4f), the quantum fluctuation operators with respect to the Fourier frequency $\omega$ can be attained. In what follows, we focus on the entanglement at $\omega=0$, and use the input-output relation $\delta A_s^{out}(\omega)=\sqrt{2\kappa}\delta A_s(\omega)-A_s^{in}(\omega)$ ($A=X,Y$) as well as the entanglement criterion $V_{ij}=\langle \delta U(\omega)\delta U^+(\omega)+\delta V(\omega)\delta V^+(\omega)\rangle<2$ proposed in Refs. [42, 43] to test the entangled feature of the cavity field outputs, where $\delta U(\omega)=\delta X_i^{out}(\omega)\pm\delta X_j^{out}(\omega)$ and $\delta V(\omega)=\delta Y_i^{out}(\omega)\mp\delta Y_j^{out}(\omega)$ ("+" in $\delta U(\omega)$ and "-" in $\delta V(\omega)$ for both $V_{0^-0^+}$ and $V_{0^+j}$, whereas "-" in $\delta U(\omega)$ and "+" in $\delta V(\omega)$ for $V_{0^-j}$). Satisfying the above inequality sufficiently demonstrates the generation of bipartite entanglement, and the smaller the correlation $V_{ij}$, the stronger the degree of the bipartite entanglement. The analytical expressions for the correlations $V_{ij}$ can easily be obtained by solving Eqs. (4a-4f); however, as they are too cumbersome to be presented here, we only give the numerical results. In the resolved sideband limit, in order to ensure the stability of the system, the coupling strength of the blue-detuned driving pump 2 field should not be larger than that of the red-detuned driving pump 1 field for the case of the equal decay rates of the cavity fields [20-23].

**Generation of Tripartite and Quadripartite Entanglements.** Figure (2a-2b) gives the main results of this study, where the behavior of tripartite and quadripartite entanglements among the outputs of the pump and probe fields as a function of the normalized effective detuning $\Delta_{eff}/\omega_m$ ($\Delta_{eff}=\Delta_0-g_0(\beta+\beta^*)$) is depicted with the realistic experimental parameters used in Ref. [29]. It can be seen from Fig. 2a that, when the pump 1 field is tuned near to the red mechanical sideband of the cavity mode 0, that is, the pump 2 field is nearly blue-detuned by $\omega_m$, and the probe 1 field is tuned to the red mechanical sideband of the neighboring cavity mode 1, the correlations $V_{0^-0^+}$, $V_{0^-1}$, and $V_{0^+1}$ all exhibit a dip in a limited range of the effective detuning around $\Delta_{eff}=\omega_m$ with the minimum values of about 1.4, 1.8, and 1.8, respectively. This clearly demonstrates that the outputs of the two pump fields and the probe field are genuinely entangled with each other, and tripartite entanglement among them is achieved. In addition, the degree of the bipartite entanglement between the two pump fields are stronger than that between the probe and either of the two pump fields.

The physical mechanism underlying the above generated tripartite entanglement among the outputs of the pump and probe fields can be well understood by considering the interaction between the cavity fields and mechanical oscillator. As seen from Eqs. (3a-3c), the effective interaction Hamiltonian of the system can be equivalently described as

$$H=\hbar\Delta_s a_s^+ a_s + \hbar\omega_m b_0^+ b_0 + \sum_s \hbar g_s a_s^+ a_s(b_0+b_0^+). \tag{5}$$

Obviously, the single cavity driven by two strong pump fields tuned, respectively, to the red and blue mechanical sidebands of a single cavity field mode in the present scenario has the similar features as two driven cavities with opposite detunings equal to the mechanical frequency coupled to a single mode of a common mechanical resonator studied in Refs. [20-23]. As analyzed in Refs. [20-23], the effective

Hamiltonian in Eq. (5) can be described by two separate two-mode interactions in the resolved-sideband regime by neglecting the counter-rotating terms. One is a nondegenerate parametric down-conversion process for the driving pump 2 field, generating a Stokes photon at the cavity resonance and a phonon at the mechanical resonance, which would result in two-mode squeezing and bipartite entanglement between the Stokes field mode and vibrational mode; and the other is a beam-splitter-like coupling between the mechanical resonator and cavity field, where an injected drive photon from pump 1 field and a phonon emitted from the vibrating mirror produce an anti-Stokes photon at the cavity resonance, and entanglement swapping leads to bipartite entanglement between the anti-Stokes field mode and vibrational mode. The above generated Stokes and anti-Stokes modes (i.e., the two components of the cavity field mode $\omega_{c0}$) can be equivalently regarded as the result of frequency down- (or up-) conversion process through scattering the two input pump fields off the common mechanical oscillator, which acts as a frequency converter with frequency equal to its oscillation frequency. Clearly, the present scheme has the similar feature as that for generating anti-Stokes-Stokes-atom tripartite entanglement via atomic spin coherence in the Λ-type atomic system studied in Refs. [8, 9], where the mechanical oscillator plays a role similar to the atomic spin coherence. Therefore, strong anti-Stokes-Stokes-mirror tripartite entanglement and subsequent tripartite entanglement among the outputs of the two pump fields and mirror can be realized.

When a relatively weak probe field with red-detuning equal to the mechanical frequency is incident onto the cavity (we choose $\Delta_j = \omega_m \gg g_j |a_j|$), its influence on the mechanical mode can be neglected; as done in Ref. [15], in the frame rotating at the mechanical frequency $\omega_m$ and under the rotating-wave approximation, the quantum fluctuation of the probe field output can be expressed as $\delta a_j^{out} = i \frac{g_j \alpha_j}{\sqrt{k}} \delta b + a_j^{in}$. It is obvious that the probe field output completely characterizes the feature of the mechanical oscillation mode, and gets entangled with the vibrating mirror and subsequent outputs of the two pump fields. In this regard, the probe field has the similar function as that used in Ref. [15] to measure the optomechanical entanglement by using an additional adjacent cavity sharing the common vibrating mirror.

The physical origin of the generated tripartite entanglement among the outputs of the two pump fields and probe field can also be seen directly from Eqs. (3a-3c). As shown in Eqs. (3a-3c), all of the pump and probe field modes are optomechanically coupled to the mechanical oscillation mode via the mechanical oscillator displacement operator $X_b$ and interact with each other, and correlation and entanglement among the cavity field modes as well as their corresponding outputs can be established. If there is no mechanical oscillator displacement, then the cavity modes would have no mutual coupling, and no correlation and entanglement would exist among them.

The above idea for producing tripartite entanglement via mechanical oscillator displacement can, in principle, be readily extended to generate any number of entangled fields when more probe fields red-detuned by the mechanical frequency are employed to drive different neighboring cavity modes. This concept is evidenced by adding the probe 2 field to test the quadripartite entanglement among the outputs of the two pump and two probe fields. As shown in Fig. 2b, all of the six correlations $V_{0^-0^+}$, $V_{0^-1}$, $V_{0^+1}$ $V_{12}$, $V_{0^-2}$, and $V_{0^+2}$ are smaller than 2 in a limited range of effective detuning around $\Delta_{eff} = \omega_m$ (note that $V_{0^-1}$ ($V_{0^+1}$) and $V_{0^-2}$ ($V_{0^+2}$) have the same evolution behavior and for clarity $V_{0^-2}$ and $V_{0^+2}$ are enlarged by 1.01 times), indicating the generation of quadripartite entanglement among these four output fields. Therefore, the realization of the quadripartite entanglement provides a clear evidence that arbitrarily multiple entangled CV fields can, in principle, be produced via the mechanical oscillator displacement induced by the two strong pump fields. The scalability to conveniently and efficiently generate arbitrary number of nondegenerate entangled fields is the key feature in the present scenario.

We also show the 3D plots of the evolution of the correlations $V_{0^-0^+}$, $V_{0^-1}$, and $V_{0^+1}$ at zero Fourier frequency with respect to the intensity ratio $R$ ($R = P_{l2}/P_{l1}$) and relative initial phase $\varphi$ of the two pump fields in Fig. (3a-3c) as well as the environmental temperature $T$ and cavity field decay rate $\kappa$ in Fig. (4a-4c) within the experimentally available parameters, respectively. It can be seen that tripartite entanglement among the outputs of the two pump fields and probe field can be realized and optimized by suitably choosing the intensity ratio $R$ of the two pump fields (nearly equal coupling strengths for the case of equal cavity decay rates), which is consistent to that in Refs. [20-23]. Also, there exists an optimal relative initial phase $\varphi$ and cavity field decay rate for generating the strongest tripartite entanglement, which indicates that by controlling the relative initial phase of the two input pump fields and/or cavity field decay rate, the generation and manipulation of the multipartite entanglement can be realized. Note that such dissipation-induced entanglement in an optomechanical sytem [17-23] and in an atomic system [36, 37, 44, 45] has been extensively examined as well. In addition, though the degree of bipartite entanglement between either of the two pump fields and the probe field would be weakened dramatically with the increase of the temperature, the bipartite entanglement between the two pump fields exhibits strong robustness to the temperature, and can still exist at room or even higher temperature.

In conclusion, we have proposed a convenient and flexible way to produce multicolor multipartite CV entanglement via mechanical oscillator displacement induced by two strong input pump fields in the traditional single-cavity optomechanical system. This method provides a proof-of-principle demonstration of efficiently and conveniently generating any number of narrow-band nondegenerate entangled fields with long correlation time, which may find potential applications in realistic quantum information processing and quantum networks.

**ACKNOWLEDGEMENTS**

This work is supported by National Natural Science Foundation of China (Nos. 12174243, and 12174288). Yang's e-mail is yangxh@shu.edu.cn, and Zhang's e-mail is junxiang_zhang@zju.edu.cn.

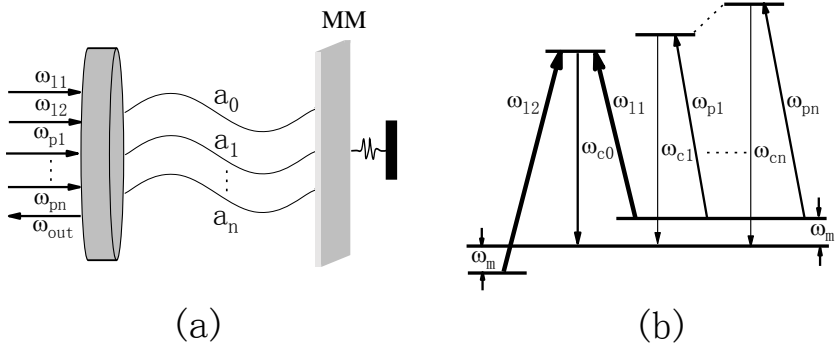

FIG. 1. (**a**) The considered conventional single-cavity optomechanical system with a movable mirror (MM) driven by two strong input pump 1 and 2 fields with frequencies $\omega_{l1}$ and $\omega_{l2}$ and any number of relatively weak input probe fields 1, 2,…n with frequencies $\omega_{p1,p2,...pn}$ and their corresponding outputs. (**b**) The relevant frequencies of the cavity fields as well as the input pump and probe fields, where the input pump 1 (2) field is red (blue)-detuned by the mechanical oscillation frequency $\omega_m$ to the cavity field mode 0 with frequency $\omega_{c0}$, and the input probe fields 1, 2,…n are tuned to the red mechanical sidebands of the neighboring cavity modes with frequencies $\omega_{c1,c2,...cn}$.

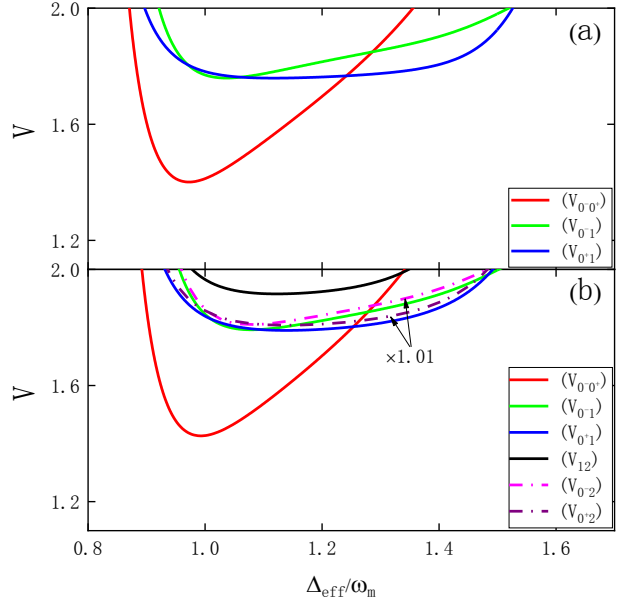

FIG. 2. The evolution of the correlations $V_{0^-0^+}$, $V_{0^-1}$, and $V_{0^+1}$ (**a**) and $V_{0^-0^+}$, $V_{0^-1}$, $V_{0^+1}$, $V_{12}$, $V_{0^-2}$, and $V_{0^+2}$ (**b**) at zero Fourier frequency as a function of the normalized effective detuning $\Delta_{eff}/\omega_m$ with the realistic experimental parameters used in Ref. [29] with $L=0.025m$, $T=0.1K$, $\omega_m = 2\pi \times 1MHz$, $\gamma_m = 2\pi \times 1Hz$, $\kappa = 2\pi \times 4.3 \times 10^5 Hz$, $\lambda = 1064nm$, $m = 150 \times 10^{-12} kg$, $P_{l1} = P_{l2} = 50 P_{p1,p2} = 40mW$, $\Delta_1 = \Delta_2 = \omega_m$, $\delta = 2\omega_m$, and $\varphi = -0.3$.

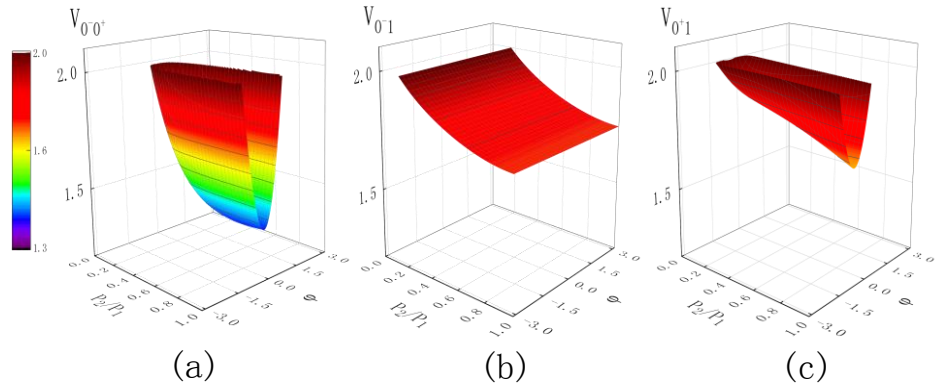

FIG. 3. The 3D plots of the evolution of the correlations $V_{0^-0^+}$ (**a**), $V_{0^-1}$ (**b**), and $V_{0^+1}$ (**c**) at zero Fourier frequency with respect to the intensity ratio $R$ ($R = P_{l2}/P_{l1}$) and relative initial phase $\varphi$ of the two driving pump fields, and the other parameters are the same as those in Fig. 2.

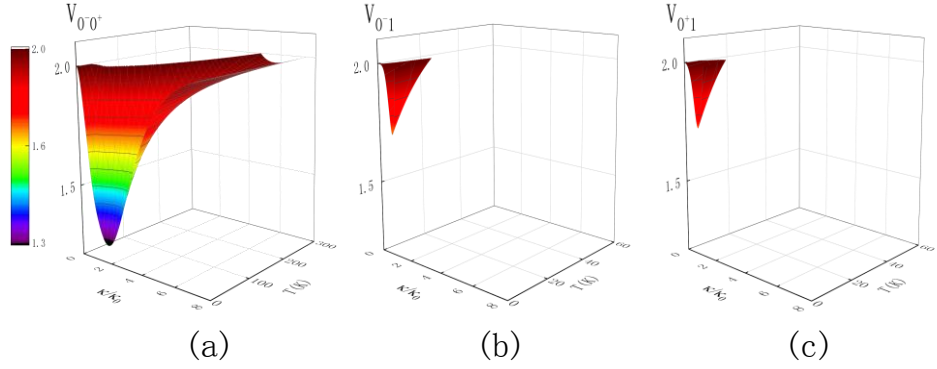

FIG. 4. The 3D plots of the evolution of the correlations $V_{0^-0^+}$ (**a**), $V_{0^-1}$ (**b**), and $V_{0^+1}$ (**c**) at zero Fourier frequency with respect to the environmental temperature $T$ and cavity field decay rate $\kappa$ ($\kappa_0 = 2\pi \times 4.3 \times 10^5 \, Hz$), and the other parameters are the same as those in Fig. 2.